\documentstyle{article}

\renewcommand{\baselinestretch}{2}

\begin{document}
\thispagestyle{empty}
\large
\title{\LARGE STRONG OSCI\-LLA\-TI\-ONS OF CU\-MU\-LANTS \\
       OF PHO\-TON
       DIST\-RI\-BU\-TI\-ON FUN\-CTI\-ON \\
       IN SLIGH\-TLY SQUEE\-ZED
        STA\-TES }
\author{\Large V.V.Dodonov, I.M.Dremin and P.G.Polynkin\\
                   Lebedev Physics Institute,\\
        Leninsky Prospect, 53, 117924, Moscow, Russia\\
and\\
V.I.Man'ko\\
Lebedev Physics Institute and University of Naples "Federico II",\\
Mostra d'Oltremare, Pad.20, 80125 Naples, Italy}
\date{}
\maketitle
\begin{abstract}
{\Large The cumulants and factorial moments of photon distribution for squeezed
and
correlated light are calculated in terms of Chebyshev, Legendre and Laguerre
polynomials. The phenomenon of strong oscillations of the ratio of the cumulant
to factorial moment is found.}
\end{abstract}
\vspace{1.5cm}

Running title: Oscillations of cumulants in squeezed states.

Keywords: cumulants and factorial moments; slightly squeezed states; strong
oscillations.

PACS Numbers: 03.65.-w;42.50.-p

\newpage
\voffset=-1.5cm
\setcounter{page}{1}
\setcounter{equation}{0}
\section{INTRODUCTION}
The coherent light [1],[2],[3] has the photon distribution function
described by the
standard Poisson distribution. The nonclassical states of light, for example
squeezed states [4],[5],[6], Schr\"odinger cat states [7],[8],[9],
correlated states [10] have the photon distribution functions which differ
essentially from poi\-sso\-ni\-an ones de\-mon\-stra\-ting ei\-ther
su\-per\-poi\-sso\-ni\-an
or sub\-poi\-sso\-ni\-an be\-ha\-vi\-our which distinguishes the
nonclassical types of
light from the coherent light considered by definition as classical one.

  One of the important differences of photon distribution function of the
nonclassical types of light from the poissonian distribution of the coherent
light is the possible strong oscillations of the photon distribution function.
For the squeezed light these oscillations have been found in [11],[12]. For the
correlated light the existence of such oscillations has been demonstrated
in [13]. For even and odd coherent states (Schr\"odinger cat states) the fast
oscillations of the photon distribution function are connected with the
absence of the states with even numbers of photons in odd cat states and the
states with odd numbers of photons in even cat states [7],[14],[15]. Such
properties of the photon distribution oscillations for the nonclassical
types of light are preserved also for the multimode electromagnetic radiation.
It was shown for two-mode squeezed light in [16],[17],[18] and for even and odd
cat states in [19],[20].

On the other hand, in high energy physics another characteristics of the
multiparticle distribution function has been widely used, namely,
factorial moments and cumulants (for the review see [21]). It was shown that
the cumulants and the functions of these quantities are very sensitive to the
details of the particle distribution demonstrating the oscillating behaviour
[22],[23],[24].
The properties of the cumulants of the integer rank for the known
distributions as poissonian and
binomial ones are well known and for these distributions the oscillatory
behaviour is absent [24].

The behaviour of the cumulants
(in particular, their oscillations) for
the squeezed states, correlated states or other nonclassical states has not
been studied till now in all details.
At the same time, for the most general photon
distribution function corresponding to squeezed and correlated one-mode light
at finite temperature the explicit expression in terms of Hermite polynomials
of two variables has been obtained in [25] and it was widely used in
[26],[27],[28],[29].

  The aim of this work is to obtain the explicit expressions for the cumulants
and factorial moments of the photon distribution function for the squeezed and
correlated light at finite temperature. We  demonstrate
that the cumulants possess the strongly oscillating behaviour in the region of
slight squeezing where the photon distribution function itself has no
oscillations. And vice versa in the region of large squeezing, where the
photon distribution function strongly oscillates, the cumulants behave
smoothly. Thus the behaviour of the cumulants may provide very sensitive method
of detecting very small squeezing and correlation phenomena due to presence
of strong oscillations.

\section{PHO\-TON DIST\-RI\-BU\-TI\-ON FUNC\-TI\-ON \newline
         AND ITS MO\-MENTS}
Since in what follows it will be necessary to use some characteristics of
one-mode squeezed light, we will briefly review the main results obtained
(see, for example [26]).

  Let us consider the most general mixed squeezed state of
one-mode light described
by the Wigner function $W(p,x)$ of the generic Gaussian form with five
real parameters, $\langle x \rangle$, $\langle p \rangle$, $\sigma_{xx}$,
$\sigma_{pp}$, $\sigma_{px}$ (first two parameters are means of position and
momentum, others  are matrix elements of the dispersion matrix for the
position and momentum),
\begin{eqnarray}
  \lefteqn{}
    &&W(p,x)=d^{-1/2}\mbox{exp}\left\{-(2d)^{-1}\left[\sigma_{xx}
    (p-\langle p \rangle )^{2}\right.\right.\nonumber\\
    &&+\left.\left.\sigma_{pp}(x-\langle x \rangle )^{2}-2\sigma_{px}(p-
    \langle p \rangle )
    (x-\langle x \rangle )\right]\right\} \hspace{0.15cm}\mbox{,}
\end{eqnarray}
where
\begin{eqnarray}
    d=\sigma_{pp}\sigma_{xx}-\sigma_{px}^{2} \nonumber
\end{eqnarray}
is determinant of the dispersion matrix.

For the photon distribution function the following formula was obtained
[25],[26]:
\begin{eqnarray}
    P_{n}=P_{0}\frac{H_{nn}^{\bf R}(y_{1},y_{2})}{n!} \hspace{0.15cm} \mbox{,}
\end{eqnarray}
where
\begin{eqnarray}
  \lefteqn{}
    &&P_{0}=\left(d+\frac{1}{2}T+\frac{1}{4}\right)^{-1/2}\nonumber\\
    &&\times \mbox{exp}\left[
    -\frac{\langle p \rangle ^{2}(2\sigma_{xx}+1)+
    \langle x \rangle ^{2}(2\sigma_{pp}
    +1)-4\sigma_{px}\langle p \rangle \langle x \rangle }{1+2T+4d}\right]
    \nonumber
\end{eqnarray}
is the probability to have no photons,
\begin{eqnarray}
    T=\sigma_{pp}+\sigma_{xx} \nonumber
\end{eqnarray}
is the trace of the dispersion matrix, $H_{nn}^{\bf R}$ -- Hermite polynomials
of two variables.

Elements of the symmetric matrix
\begin{eqnarray}
    \bf R=\left\|\begin{array}{cc}
    R_{11} & R_{12} \\
    R_{12} & R_{22}
    \end{array} \right\| \nonumber
\end{eqnarray}
determining the Hermite polynomials are given by the formulas:
\begin{eqnarray}
    R_{11}=R_{22}^{*}=\frac{2(\sigma_{pp}-\sigma_{xx}-2i\sigma_{px})}
                           {1+2T+4d}\hspace{0.15cm} \mbox{,}\hspace{1cm}
    R_{12}=\frac{1-4d}{1+2T+4d} \hspace{0.15cm} \mbox{,} \nonumber
\end{eqnarray}
and two arguments of the polynomials are defined by the equation:
\begin{eqnarray}
    y_{1}=y_{2}^{*}=\frac{2\left[(T-1)z^{*}+(\sigma_{pp}-\sigma_{xx}+2i
    \sigma_{px})z\right]}{2T-4d-1} \hspace{0.15cm} \mbox{.} \nonumber
\end{eqnarray}
The complex parameter $z$ is given by the relation:
\begin{eqnarray}
    z=2^{-1/2}\left(\langle x \rangle +i \langle p \rangle \right)
    \hspace{0.15cm} \mbox{.}
\end{eqnarray}
The generating function for the photon distribution function was also
obtained in [26]:
\begin{eqnarray}
    G(u)=P_{0}\left[\left(1-\frac{u}{\lambda_{1}}\right)
    \left(1-\frac{u}{\lambda_{2}}\right)\right]^{-1/2}
    \mbox{exp}\left[\frac{u\xi_{1}}{u-\lambda_{1}}+\frac{u\xi_{2}}
    {u-\lambda_{2}}\right] \hspace{0.15cm} \mbox{,}
\end{eqnarray}
where
\begin{eqnarray}
  \lefteqn{}
    &&\lambda_{1}=\left(\sqrt{R_{11}R_{22}}-R_{12}\right)^{-1} \hspace{0.15cm}
    \mbox{,}
    \hspace{1cm} \lambda_{2}=-\left(\sqrt{R_{11}R_{22}}+R_{12}\right)^{-1}
    \hspace{0.15cm} \mbox{,}\nonumber\\
    &&\xi_{1}=\frac{1}{4}\left(1-\frac{R_{12}}{\sqrt{R_{11}R_{22}}}\right)
    \left(y_{1}^{2}R_{11}+y_{2}^{2}R_{22}-2\sqrt{R_{11}R_{22}}y_{1}y_{2}
    \right) \hspace{0.15cm} \mbox{,} \nonumber\\
    &&\xi_{2}=\frac{1}{4}\left(1+\frac{R_{12}}{\sqrt{R_{11}R_{22}}}\right)
    \left(y_{1}^{2}R_{11}+y_{2}^{2}R_{22}+2\sqrt{R_{11}R_{22}}y_{1}y_{2}
    \right) \hspace{0.15cm} \mbox{.} \nonumber
\end{eqnarray}
The photon distribution function is related to $G(u)$ as follows:
\begin{eqnarray}
    \left. P_{n}=\frac{1}{n!}\frac{d^{n}G(u)}{du^{n}}\right|_{u=0}
    \hspace{0.15cm} \mbox{.}
    \nonumber\\
\end{eqnarray}
The above-mentioned normalized cumulants and factorial moments are defined as
\begin{eqnarray}
  \lefteqn{}
    && \left. K_{q}=\frac{1}{\langle n \rangle ^{q}}\frac{d^{q}
    \mbox{ln}G(u)}{du^{q}}
    \right|_{u=1} \hspace{0.15cm} \mbox{,}\nonumber\\
    && \left. F_{q}=\frac{1}{\langle n \rangle ^{q}}\frac{d^{q}G(u)}
    {du^{q}}\right|_{u=1}
    \hspace{0.15cm} \mbox{,}
\end{eqnarray}
respectively, and related by the following recursion relation:
\begin{eqnarray}
    F_{q}=\sum_{m=0}^{q-1} C_{q-1}^{m}K_{q-m}F_{m} \hspace{0.15cm} \mbox{,}
\end{eqnarray}
where
\begin{eqnarray}
    C_{q-1}^{m}=\frac{(q-1)!}{m!(q-m-1)!} \nonumber
\end{eqnarray}
are the binomial coefficients. Some formulas for the cumulants of photon
distribution function can be found in [30],[31].
The factorial moments have been widely used in particle physics [21] to analyze
intermittency properties [32] of fluctuations.
However, they are less instructive than the cumulants or their functions.
It was shown in [22],[24] that the ratio of cumulant to factorial moments,
i.e. the function
\begin{eqnarray}
    H_{q}=K_{q}/F_{q} \hspace{0.15cm} \mbox{,}
\end{eqnarray}
is a very sensitive measure of tiny details of the multiplicity distribution.
In particular, it can be used to distinguish between different distributions
which otherwise look quite similar.

\section{ANALYSIS OF $K_{q}$, $F_{q}$ AND $H_{q}$}
\setcounter{equation}{0}

  It was already mentioned in the Introduction that the photon distribution
function exhibits an oscillatory behaviour if we deal with highly squeezed
states (\hspace{0.05cm} $T=\sigma_{pp}+\sigma_{xx}\gg 1$ \hspace{0.05cm} )
for large values
of the parameter $z$ (2.3).
A question arises: is it possible to obtain a similar "abnormal"
behaviour of other characteristics of the photon distribution, namely,
cumulants, factorial moments and the function $H_{q}$ defined in (2.8)?
If yes, then in what region of parameters of the function (2.1) such anomalies
can take place? The present section is dedicated to the solution of this
problem.

  The direct differentiation of the function $\mbox{ln}G(u)$ at $u=1$ yields:
\begin{eqnarray}
    K_{q}=\frac{(q-1)!}{\langle n \rangle^{q}}\left[\frac{1}
    {(\lambda_{1}-1)^{q}}
    \left(\frac{1}{2}+q\frac{\xi_{1}\lambda_{1}}{1-\lambda_{1}}\right)
    +\frac{1}{(\lambda_{2}-1)^{q}}
    \left(\frac{1}{2}+q\frac{\xi_{2}\lambda_{2}}{1-\lambda_{2}}\right)
    \right] \hspace{0.15cm} \mbox{,}
\end{eqnarray}
where [26]
\begin{eqnarray}
    \langle n \rangle=\frac{T-1}{2}+|z|^{2} \hspace{0.15cm}
    \mbox{.} \nonumber
\end{eqnarray}
High oscillation are obtained at strong squeezing (large $T$) or at
large values of $\vert z\vert ^2$.

It is known [27] that
\begin{eqnarray}
    T\geq 1 \hspace{0.15cm} \mbox{,} \hspace{1cm}
    d\geq \frac{1}{4} \hspace{0.15cm} \mbox{.}
\end{eqnarray}
Using (3.2) the following inequalities can be easily obtained:
\begin{eqnarray}
  \lefteqn{}
  &&\lambda_{1}>1 \hspace{1cm}\mbox{and} \nonumber\\
  &&\lambda_{2}<0 \hspace{1cm}\mbox{or} \hspace{1cm} \lambda_{2}>1
  \hspace{0.15cm} \mbox{.}
\end{eqnarray}
Expression in the square brackets in (3.1) consists of two terms:
\begin{eqnarray}
    \frac{1}{(\lambda_{1}-1)^{q}}
    \left(\frac{1}{2}+q\frac{\xi_{1}\lambda_{1}}{1-\lambda_{1}}\right)
    \hspace{0.15cm} \mbox{,}
\end{eqnarray}
\begin{eqnarray}
    \frac{1}{(\lambda_{2}-1)^{q}}
    \left(\frac{1}{2}+q\frac{\xi_{2}\lambda_{2}}{1-\lambda_{2}}\right)
     \hspace{0.15cm} \mbox{.}
\end{eqnarray}
The first term is of constant sign. The second one is oscillating in
the case $\lambda_{2}<0$. With aim to obtain the oscillations of the whole
function $K_{q}$ we will treat only this case:
\begin{eqnarray}
    \lambda_{2}<0 \hspace{0.15cm} \mbox{.} \nonumber
\end{eqnarray}
Then
\begin{eqnarray}
    \frac{1}{(\lambda_{2}-1)^{q}}
    =\frac{(-1)^{q}}{(1+|\lambda_{2}|)^{q}} \hspace{0.15cm} \mbox{.}
    \nonumber
\end{eqnarray}
However, then it follows that
\begin{eqnarray}
    |\lambda_{2}|\geq \lambda_{1}>1 \hspace{0.15cm} \mbox{,} \nonumber
\end{eqnarray}
and the alternating term diminishes faster than the constant sign term.
The terms $1/|\lambda_{2}-1|$ and
$1/(\lambda_{1}-1)$ are most close to one another if
\begin{eqnarray}
    d=\frac{1}{4} \hspace{1cm} \mbox{(the pure state),} \nonumber
\end{eqnarray}
that is used in the following.

  First of all we consider the simplest case when the value of
$z$ as given by (2.3) equals to zero. Then
\begin{eqnarray}
  \lefteqn{}
    &&K_{q}=\frac{\beta^{q/2}}{\langle n \rangle ^{q}}(q-1)!
    T_{q}(\alpha) \hspace{0.15cm} \mbox{,}\nonumber\\
    &&F_{q}=\frac{\beta^{q/2}}{\langle n \rangle ^{q}}q!P_{q}(\alpha)
    \hspace{0.15cm} \mbox{,}
\end{eqnarray}
where
\begin{eqnarray}
    \beta=d+\frac{1}{4}-\frac{T}{2} \hspace{0.15cm} \mbox{,} \hspace{1cm}
    \alpha=\frac{T-1}{\sqrt{4d+1-2T}} \hspace{0.15cm} \mbox{,}\nonumber
\end{eqnarray}
$T_{q}(\alpha)$ and $P_{q}(\alpha)$ are the Chebyshev polynomials of
the first kind and Legendre polynomials, respectively.
Let us note that the arguments of polynomials are purely imaginary
but the whole expressions for moments are real, surely.

  For $H_{q}$ we obtain the expression:
\begin{eqnarray}
    H_{q}=\frac{T_{q}(\alpha)}{qP_{q}(\alpha)} \hspace{0.15cm} \mbox{.}
\end{eqnarray}
  Taking $d=1/4$ we have:
\begin{eqnarray}
  \lefteqn{}
    &&K_{q}=\left(\frac{2}{1-T}\right)^{q/2}(q-1)!T_{q}(\alpha)
    \hspace{0.15cm} \mbox{,}
    \nonumber\\
    &&F_{q}=\left(\frac{2}{1-T}\right)^{q/2}q!P_{q}(\alpha)
    \hspace{0.15cm} \mbox{,}
    \nonumber\\
    &&\alpha=\sqrt{\frac{1-T}{2}} \hspace{0.15cm} \mbox{.} \nonumber
\end{eqnarray}

  In this case the curve $H_{q}$ has step-like shape at $(T-1)\rightarrow 0$;
steps become smoothed as $T$ grows (fig.$1$). We should note that direct
limit $T\rightarrow 1$ shows the discontinuous character
of the function $H_{q}(T)$ at $T=1$. The point is that at $T=1$ we are dealing
with the usual Poisson distribution (let us remind that we treat a case
$d=1/4$,
$|z|=0$), where $H_{q}=\delta_{q1}$, i.e. $H_{1}=1$, $H_{q}=0$ at $q\neq 1$,
which differs from the behaviour of $H_{q}$
at $(T-1)=10^{-5}$, depicted in the fig $1$.
The particular values of the second rank moments are very high what
reveals extremely wide distribution (so wide distributions are
unknown in particle multiproduction, for example). Namely, one can
easily show that $F_2>3, K_2>2, H_2>2/3$.

  Consider now the case $|z|\neq 0$. Since the photon distribution function
is invariant with respect to rotation in a phase space, without loss of
generality we can consider $\sigma_{xx}=\sigma_{pp}$ ($\sigma_{px}\neq 0$ --
correlated state). By appropriate choice of the phase of (2.3)
($\langle x \rangle =-\langle p \rangle $) we
cancel the linearly growing term $q\xi_{1}\lambda_{1}/(1-\lambda_{1})$ in
(3.4).
Moreover, the analogous linear term $q\xi_{2}\lambda_{2}/(1-\lambda_{2})$
in (3.5) becomes maximal at fixed $|z|$. Thus we have left only two
variable parameters $T$ and $|z|$, and formula (3.1) has the following
final form:
\begin{eqnarray}
    K_{q}=\frac{(q-1)!}{\left(\frac{T-1}{2}+|z|^{2}\right)^{q}}
    \left[\frac{1}{2(\lambda_{1}-1)^{q}}+\frac{(-1)^{q}}{(1+|\lambda_{2}|)^{q}}
    \left(\frac{1}{2}-q\frac{\xi_{2}|\lambda_{2}|}{1+|\lambda_{2}|}\right)
    \right] \hspace{0.15cm} \mbox{,}
\end{eqnarray}
where
\begin{eqnarray}
   \lefteqn{}
     &&\lambda_{1}=-\lambda_{2}=\sqrt{\frac{T+1}{T-1}} \hspace{0.15cm} \mbox{,}
     \nonumber\\
     &&\xi_{2}=2\left(\frac{T}{\sqrt{T^{2}-1}}+1\right)|z|^{2}
     \hspace{0.15cm} \mbox{.}
     \nonumber
\end{eqnarray}

  In the case of large $T$ (highly squeezed state) we can obtain the finite
number of oscillations of $K_{q}$ taking large value of $|z|$.
However, the average number of photons in corresponding states is large, and
the amplitude of the oscillations decreases exponentially due to
the factor $1/\langle n \rangle ^{q}$. Remind that in this very case the strong
oscillations
of the photon distribution function can be observed.

  Now let us consider the case of the slightly squeezed state, $y=(T-1)\ll 1$,
when photon
distribution function does not oscillate. Impose also an additional condition
\begin{eqnarray}
    \gamma=\frac{|z|^{2}}{\sqrt{y/2}}\gg 1 \hspace{0.15cm} \mbox{,} \nonumber
\end{eqnarray}
that makes possible to obtain approximate formulas for the functions $K_{q}$,
$F_{q}$ and $H_{q}$.
For $K_{q}$ we have the following approximate expression:
\begin{eqnarray}
    K_{q}=q!(-1)^{q-1}\gamma^{1-q} \hspace{0.15cm} \mbox{.}
\end{eqnarray}
Then recursion relation (2.7) yields:
\begin{eqnarray}
    F_{q}=q!(-1)^{q}\gamma^{-q}L_{q}^{-1}(\gamma) \hspace{0.15cm} \mbox{,}
\end{eqnarray}
where $L_{q}^{-1}(x)$ are generalized Laguerre polynomials. For $H_{q}$
with $ q \ll \gamma $ we have:
\begin{eqnarray}
    H_{q}=K_{q}/F_{q}=-\frac{\gamma}{L_{q}^{-1}(\gamma)}\approx (-1)^{q+1}
    q!\gamma^{1-q} \ll 1 \hspace{0.15cm} \mbox{.}
\end{eqnarray}
(If $\gamma\gg q$, the term with the highest power of $\gamma$ dominates
over the rest of the sum in $L_{q}^{-1}(\gamma)$, and $F_{q} \rightarrow 1$
as for Poisson distribution). The exact shape of the function
$H_{q}$ is shown in fig.$2$.
The distribution function $P_{n}$ does not oscillate (fig.$2^{a}$).

  However, the most abrupt oscillations of the functions $K_{q}$ and $H_{q}$
have been obtained when $(T-1)\ll 1$, but condition $\gamma\gg 1$
is not valid. The
corresponding curves are shown in the figs.$3$, $3^{a}$. Note that the photon
distribution function is smooth again being approximately equal to zero at
$q\neq 1$.

  The most regular oscillating patterns of $K_{q}$ and $H_{q}$ are seen at
$(T-1)\sim 0.1$, $|z|\sim 1$ (figs.$4$, $4^{a}$).

  The alternating sign cumulants are typical also for the fixed multiplicity
distribution, i.e., for $P_{n}=\delta_{nn_{0}}$ ($n_{0}=const$) [24].
Let us note that there exist smooth multiplicity distributions which give rise
to cumulants oscillating with larger period (see [22]).

  Finally we consider the opposite case when the photon
distribution function
$P_{n}$ exhibits strong oscillations while $K_{q}$ and $H_{q}$
behave smoothly.
Such a behaviour is typical at $T\sim 100$, $|z|\sim 1$  when $K_{q}$
exponentially grows while $H_{q}$ monotonically decreases with $q$ (fig.5).

\section{CONCLUSION}

   We have shown  that cumulants of the photon distribution
function and their ratio to the corresponding factorial moments exhibit
oscillating behaviour in the case of slightly squeezed states:
$(T-1)=(\sigma_{pp}+\sigma_{qq})-1\ll 1$. We have considered also the pure
state with $d=1/4$. Oscillations of the photon distribution function
are absent in that case.

  We are yet unable to establish direct correspondence between the oscillations
of cumulants and the behaviour of the photon distribution function. Somehow it
should depend on the range of parameters considered. For example, it is known
[26] that parameter
$d=\sigma_{pp}\sigma_{xx}-\sigma_{px}^{2}$ characterizes the temperature of the
system and the temperature equals to zero at $d=1/4$.
Is it possible to observe oscillations of cumulants and of the function
$H_{q}$ at other parameters of the function (2.1), for example, at
$d>1/4$? This problem has not been studied in all details yet. However,
evidently, the oscillations must be smoothed as $d$ grows since it is
analogous to the behaviour of the function $P_{n}$,
whose oscillations disappear
with increase of temperature. The extcusion of the above procedure to
non-integer ranks is straightforward.
We hope to consider these problems in subsequent publications.

\section*{ACKNOWLEDGEMENTS}

  This work was supported by Russian State Science and Technology Program
"Fundamental Nuclear Physics". One of us (V.I.M) thanks the University of
Naples "Federico II" for kind hospitality.

\newpage
\begin{center}
    {\Large FIGURE CAPTIONS}
\end{center}
\renewcommand{\baselinestretch}{1}

Figure $1$: The behaviour of the function $H_{q}$ defined in (2.8) at
$d=1/4$, $|z|=0$; parameter $T$ is varied: (1) $(T-1)=10^{-5}$, (2) $T=1.1$,
(3) $T=1.2$.

Figure $2$: The behaviour of the function $H_{q}$ at $d=1/4$, $T=1.01$;
the curves (1) and (2) correspond to the values
$|z|^{2}=1.01, \hspace{0.1cm} 0.8$, respectively.

Figure $2^{a}$: The photon distribution function at $d=1/4$, $T=1.01$;
the curves in order of the lowering maxima correspond respectively to the
values $|z|^{2}=1.01, \hspace{0.1cm} 0.8$.

Figure $3$: The cumulants of the photon distribution function at $d=1/4$,
$(T-1)=10^{-5}$, $|z|^{2}=0.01$.

Figure $3^{a}$: The behaviour of the function $H_{q}$ at $d=1/4$,
$(T-1)=10^{-5}$; parameter $|z|$ is varied:
(1) $|z|^{2}=0.01$, (2) $|z|^{2}=0.005$, (3) $|z|^{2}=0.01$.

Figure $4$: The cumulants of the photon distribution function at $d=1/4$,
$T=1.1$, $|z|^{2}=1.1$.

Figure $4^{a}$: The behaviour of the function $H_{q}$ at $d=1/4$, $T=1.1$;
parameter $|z|$ is varied: (1) $|z|^{2}=2$, (2) $|z|^{2}=1.1$.

Figure $4^{b}$: The photon distribution function at $d=1/4$, $T=1.1$;
the curves in order of the lowering maxima correspond to the
values $|z|^{2}=2, \hspace{0.1cm} 1.1$, respectively.

Figure $5$: The smooth curve for the function $H_{q}$ and the oscillating
photon distribution function $P_{n}$ at $d=1/4$, $T=100$, $|z|=1$.

\end{document}